\documentclass[aps,prl,twocolumn,footinbib,showpacs,floatfix]{revtex4}

\usepackage{amssymb}
\usepackage[pdftex]{graphicx}	
	\DeclareGraphicsExtensions{.pdf,.png,.jpg}
\usepackage[applemac]{inputenc}
\usepackage{amsmath,amssymb,amsthm}
\usepackage{mathbbol,bbm}
\usepackage{graphicx,float}
\usepackage[final]{showkeys}
\usepackage{bm,dsfont}

\def\ket#1{\left|#1\right>}

\def\Tr{ {\rm{Tr }}}

\usepackage{booktabs}

\begin{document}
\title{Two-Dimensional Density-Matrix Topological Fermionic Phases: \\
Topological Uhlmann Numbers}
\author{O. Viyuela, A. Rivas and M. A. Martin-Delgado}
\affiliation{Departamento de F\'{\i}sica Te\'orica I, Universidad Complutense, 28040 Madrid, Spain}

\vspace{-3.5cm}

\begin{abstract}
We construct a topological invariant that classifies density matrices of symmetry-protected topological orders in two-dimensional fermionic systems. As it is constructed out of the previously introduced Uhlmann phase, we refer to it as the topological Uhlmann number ${\rm n}_{\rm U}$. With it, we study thermal topological phases in several two-dimensional models of topological insulators and superconductors, computing phase diagrams where the temperature $T$ is on an equal footing with the coupling constants in the Hamiltonian. Moreover, we find novel thermal-topological transitions between two nontrivial phases in a model with high Chern numbers. At small temperatures we recover the standard topological phases as the Uhlmann number approaches to the Chern number.
\end{abstract}


\maketitle

\noindent {\it 1. Introduction.---}
Intrinsic topological orders (TOs) \cite{wenbook04,wen89,wen_niu90,ASW84}, and symmetry-protected topological orders (SP-TOs), like
topological insulators and superconductors in fermionic \cite{Haldane_88,kane_mele05,bernevig_zhang06,kane_mele05b,moore_balents07,fu_kane_mele07,rmp1,rmp2,LibroBernevig}, or more recently, bosonic \cite{VS13,MKF13,Senthil14,YW13,wen_etal14} systems, have been extensively studied and classified. Although these studies provide a successful picture for quantum systems in pure states, typically the ground state, very little is known about the fate of those topological phases of matter when the system is in a mixed quantum state represented by a density matrix. In fact, the correct understanding of this situation becomes particularly relevant in order to address unavoidable thermal effects on topological phases and nonequilibrium dynamics under dissipation \cite{Bardyn_et_al13}.

Over the last few years, there have been some results establishing the absence of stable topological phases subject to thermal effects, like for TOs with spins \cite{AFH,Hastings,Selfcorrecting}, for SP-TOs with spins \cite{Evert}, or SP-TOs with fermions under certain conditions \cite{Viyuela_et_al12}. However, in a recent work \cite{Viyuela_et_al14} we have shown that it is possible to characterize a thermal
topological phase for topological insulators and superconductors in one-dimensional systems. This thermal topological phase, classified by Uhlmann holonomies \cite{Uhlmann86,Uhlmann89}, is separated from a trivial phase by a critical finite temperature, at which the topological phase abruptly disappears. This result paves the way towards the characterization of SP-TOs with fermions in thermal states, or more general density matrices.

In this paper, we have achieved this goal for two-dimensional fermion systems, either insulating or superconducting. This extension from one to two-dimensional systems is nontrivial in the sense that a direct generalization of the topological invariants for pure states (Chern numbers) to density matrices via the Uhlmann approach leads to trivial results. However, we have succeeded in circumventing this problem and introducing a suitable notion of topological Uhlmann numbers. These are gauge invariant and observable quantities which allows for a classification of topological phases of density matrices in two-dimensional (2D) quantum systems. Specifically, we have applied this approach to determine new topological phase diagrams including temperature for three emblematic models of 2D insulators and superconductors. As a result, we have found a thermal topological phase for a 2D chiral \textit{p}-wave superconductor, which can host vortices with non-Abelian Majorana fermions, and thermal-topological transitions between two nontrivial phases in a model with high Chern numbers.

First of all, let us briefly recall the basic concepts of the Uhlmann approach, the reader may see \cite{Viyuela_et_al14,Uhlmann86,Uhlmann89} for a more detailed picture. Let $\mathcal{Q}$ denote the convex set of density matrices. For some $\rho\in \mathcal{Q}$, any of the matrices $w$ such that $\rho=ww^\dagger$ is called an amplitude of $\rho$. The set of amplitudes generates the set $\mathcal{Q}$ via this equation and forms a Hilbert space $\mathcal{H}_w$ with the Hilbert-Schmidt product $(w_1,w_2):=\Tr(w_1^\dagger w_2)$. This aims to be the density-matrix analogy to the standard situation where vector states $\ket{\psi}$ span a Hilbert space and generate pure states by the relation $|\psi\rangle\langle\psi|$. Actually, the phase freedom of pure states [U(1)-gauge freedom], is generalized to a ${\rm U}(n)$-gauge freedom ($n$ is the dimension of the space), as $w$ and $wU$  are amplitudes of the same density matrix for some unitary operator $U$.

Now, let $\bm{k}(t)|_{t=0}^1$ define a (closed) trajectory along a family of density matrices parametrized by $\bm{k}$, $\rho_{\bm{k}}$. By defining a proper parallel transport condition on the amplitudes $w_{\bm{k}(t)}$, $\rho_{\bm{k}(t)}=w_{\bm{k}(t)}w_{\bm{k}(t)}^\dagger$, it is possible to define a geometric phase for density matrices via the associated holonomy. More concretely, after the parallel transportation we have $w_{\bm{k}(1)}=w_{\bm{k}(0)}V$, with unitary $V=\mathcal{P}{\rm e}^{\oint A_{\rm U}}$; where $\mathcal{P}$ stands for the path ordering operator along the trajectory $\bm{k}(t)|_{t=0}^1$, and $A_{\rm U}=\sum_\mu A^{\rm U}_\mu(\bm{k}) dk_\mu$ is the so-called Uhlmann connection form. The geometric phase is defined from the mismatch between the initial point $w_{\bm{k}(0)}$ and final point $w_{\bm{k}(1)}=w_{\bm{k}(0)}V$. Specifically,
\begin{equation}
\Phi_{\rm U}:=\arg(w_{\bm{k}(0)},w_{\bm{k}(1)}),
\end{equation}
which is a gauge-independent quantity \cite{Uhlmann86,Uhlmann89}. In the particular gauge where $w_{\bm{k}(0)}=\sqrt{\rho_{\bm{k}(0)}}$, it takes the simple form
\begin{equation}\label{phiU}
\Phi_{\rm U}=\arg(w_{\bm{k}(0)},w_{\bm{k}(1)})=\arg \Tr\left[\rho_{\bm{k}(0)} \mathcal{P}{\rm e}^{\oint A_{\rm U}}\right],
\end{equation}
where the components of the connection are given by \cite{Hubner93}
\begin{equation}\label{dAfinal}
A_{\rm U}=\sum_{\mu,i,j}|\psi^i_{\bm{k}}\rangle\frac{\langle\psi^i_{\bm{k}}|\left[(\partial_\mu\sqrt{\rho_{\bm{k}}}),\sqrt{\rho_{\bm{k}}}\right]|\psi^j_{\bm{k}}\rangle}{p^i_{\bm{k}}+p^j_{\bm{k}}}\langle\psi^j_{\bm{k}}|dk_\mu,
\end{equation}
in the spectral basis of $\rho_{\bm{k}}=\sum_jp^j_{\bm{k}}|\psi^j_{\bm{k}}\rangle\langle\psi^j_{\bm{k}}|$, with $\partial_\mu:=\frac{\partial}{\partial k_\mu}$.

The phase $\Phi_{\rm U}$, which is experimentally observable \cite{Ericsson2003,Sjoqvist_07,Zhu_et_al11}, has a purely geometric meaning in the sense that it depends only on the geometry of the trajectory. Although in general $\Phi_{\rm U}$ may change with the starting point of the trajectory $\bm{k}(0)$, as we shall see, it can be used to construct topological invariants that are independent of this starting point.

At zero temperature, the standard method to define topological invariants in two-dimensional SP-TO systems is by means of Chern numbers. In the simplest scenario, we consider a time-reversal broken two-band system with the Fermi energy between both bands. Then the Chern number is given by
\begin{equation}\label{ChernN}
{\rm Ch}:=\frac{1}{2\pi}\int_{\rm BZ}d^2\bm{k} F_{xy}(\bm{k}),\quad F_{xy}(\bm{k}):=\partial_{x}A_{y}(\bm{k})-\partial_{y}A_{x}(\bm{k}),
\end{equation}
where BZ stands for Brillouin zone, $A_j(\bm{k})={\rm i} \langle u_{\bm{k}}|\partial_{j} u_{\bm{k}}\rangle$ is the Berry connection and $|u_{\bm{k}}\rangle$ is the eigenvector corresponding to the lower energy band. This number is a topological invariant which only takes on integer values. While this kind of constructions can be extended to higher dimensional systems or systems with time-reversal symmetry \cite{LibroBernevig}, when attempting the generalization to density matrices via the Uhlmann connection, one finds the following  fundamental obstruction.

\smallskip

\noindent {\it 2. Triviality of the Uhlmann Chern number.---}
The natural way to generalize the Chern number to arbitrary density matrices is to consider the first Chern class associated to the Uhlmann curvature, which is constructed from the Uhlmann connection $A_{\rm U}$ via the standard formula for the non-Abelian case, $F_{xy}^{\rm U}=\partial_{x}A^{\rm U}_{y}-\partial_{y}A^{\rm U}_{x}+[A^{\rm U}_{x},A^{\rm U}_{y}]$. Then, according to the theory of characteristics classes \cite{Eguchi,Nakahara}, the (first) Chern number of the Uhlmann curvature would be given by ${\rm Ch}_{\rm U}:=\tfrac{{\rm i}}{2\pi}\int_{\rm BZ}d^2\bm{k} \Tr (F^{\rm U}_{xy})$; however, this number turns out to be always zero. The reason for this is twofold: on the one hand, the Uhlmann connection belongs to the $\mathfrak{su}(n)$ Lie algebra, so its trace vanishes and so does the trace of its curvature; on the other hand, the Chern number is 0 if there is a smooth gauge defined along the whole BZ \cite{Eguchi,Nakahara}, and this is the case for the Uhlmann U(n) gauge. We can take the gauge $w_{\bm k}=\sqrt{\rho_{\bm{k}}}$ which is well defined provided that
$\rho_{\bm{k}}$ is not singular at some crystalline momentum $\bm{k}$, which is a rather natural condition \cite{footnote}. Therefore ${\rm Ch}_{\rm U}=0$ in any case.

This makes not obvious the extension of 2D topological invariants by means of the Uhlmann approach. We hereby show the way to circumvent this obstruction.

\begin{figure}[t]
	\includegraphics[width=\columnwidth]{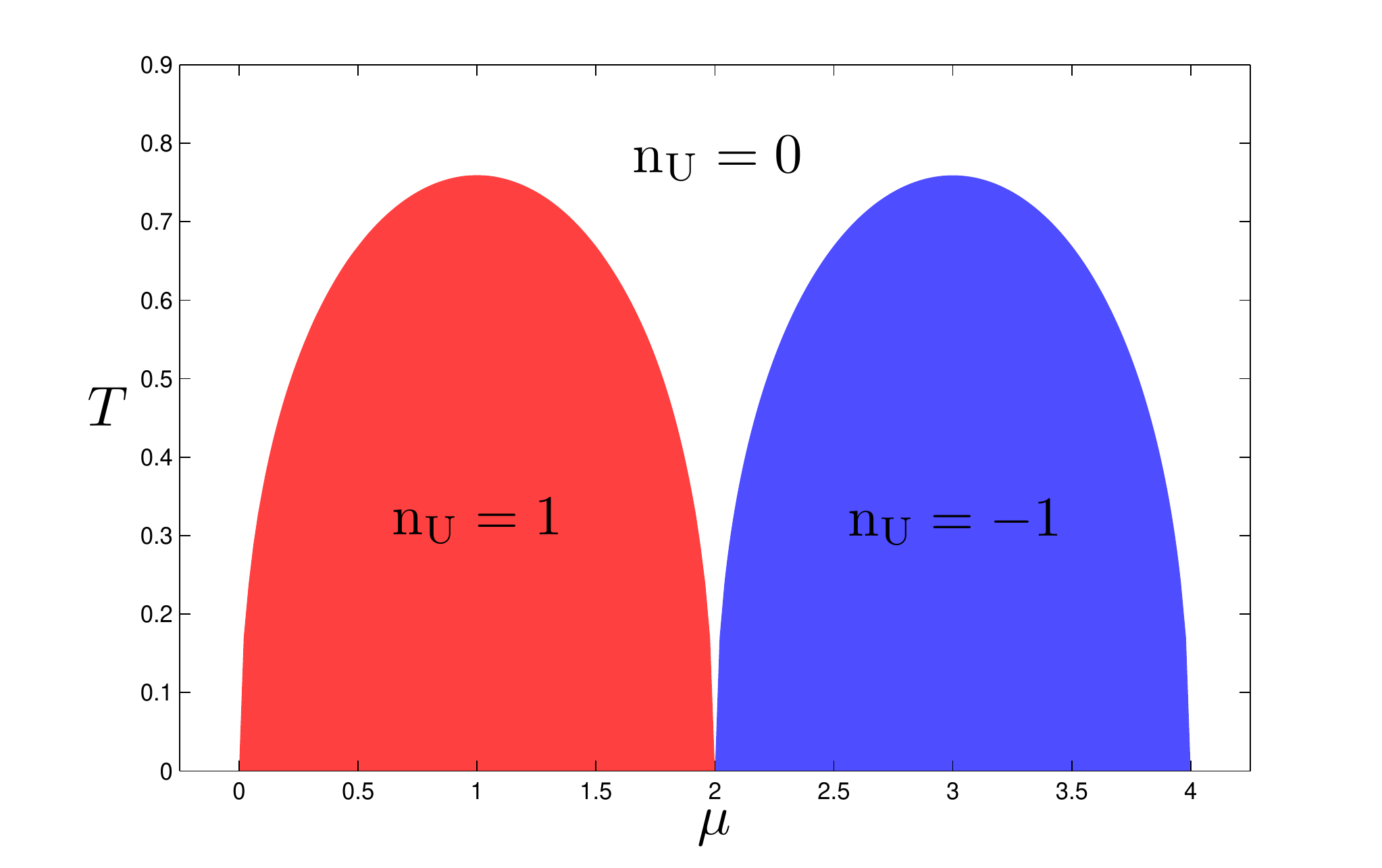}
	\caption{Topological Uhlmann phases for the \textit{p}-wave superconductor model as a function of the chemical potential $\mu$ and the temperature $T$. At $T=0$, the system has ${\rm Ch}=+1$ for $0<\mu<2$, ${\rm Ch}=-1$ for $2<\mu<4$, and ${\rm Ch}=0$ otherwise. As $T$ increases, the nontrivial phases remain up to some critical temperature $T_c$ at which Uhlmann number goes from ${\rm n}_{\rm U}=\pm1$ to ${\rm n}_{\rm U}=0$.}
	\label{Fig1}
\end{figure}

\smallskip

\noindent {\it 3. Topological Uhlmann Numbers.---}
The fact that ${\rm Ch}_{\rm U}$ becomes identically zero does not mean that all topological properties of density matrices are trivial. If this assertion were true, we could not claim that systems at $T=0$ display topological order, as they are just a particular case of generally mixed density matrices. Note that ${\rm Ch}_{\rm U}$ is not the only topological invariant that we can construct on a torus. Actually, in the Berry case, the Chern number \eqref{ChernN} can be rewritten as \cite{Abanin_et_al_13,Resta,Xiao}
\begin{equation} \label{ChernW}
{\rm Ch}=\frac{1}{2\pi}\oint dk_x \frac{d\Phi_{\rm B}(k_x)}{d k_x},
\end{equation}
where $\Phi_{\rm B}(k_x)=\oint dk_y A_y(k_x,k_y)$ is the Berry phase along the $k_y$-nontrivial homological circle of the torus at the point $k_x$, and $\oint dk_x$ denotes the integration along the $k_x$-nontrivial homological circle. To prove the equality \eqref{ChernW}, one divides the surface integral \eqref{ChernN} in small slices along the $k_x$ direction and applies the Stokes' theorem to each of them [the U(1) gauge, which may be ill defined over the whole BZ, is always well defined in a sufficiently small slice]. Then, in the limit of slices with infinitesimal width, the sum becomes an integral and one immediately obtains Eq. \eqref{ChernW}. If $\Phi_{\rm B}(k_x)$ displays some $2\pi$-discontinuous jumps along the $k_x$ circle, we take $\Phi_{\rm B}(k_x)$ to be a smooth function equal to $\oint dk_y A_y(k_x,k_y)$ mod. $2\pi$ in to order to calculate its derivative. Actually, what the Chern number is measuring is the number of those $2\pi$ jumps, i.e. the number of windings of $\Phi_{\rm B}(k_x)$ as the $k_x$ circle is covered. This is clearly a topological invariant, particularly a winding number. It associates every state of the system with an homotopy class of the Berry phase mapping $\Phi_{\rm B}(k_x):S^1\rightarrow S^1$, between the nontrivial homological circle $S^1$ and the complex phases $U(1)\cong S^1$.

Remarkably, in contrast to Eq. \eqref{ChernN}, the equivalent formula \eqref{ChernW} allows for a nontrivial generalization to density matrices. To that aim, we proceed by substituting the Berry phase $\Phi_{\rm B}(k_x)$ by the Uhlmann phase $\Phi_{\rm U}(k_x)$, Eq. \eqref{phiU}, in Eq. \eqref{ChernW}:
\begin{equation}\label{nU}
{\rm n}_{\rm U}:=\frac{1}{2\pi}\oint dk_x \frac{d\Phi_{\rm U}(k_x)}{d k_x}.
\end{equation}
Analogously to the Berry case, this integer number is a topological invariant which classifies the density matrices of a quantum system according to the homotopy class of the Uhlmann phase mapping, $\Phi_{\rm U}(k_x):S^1\rightarrow S^1$. Moreover, since for pure states $\Phi_{\rm U}=\Phi_{\rm B}$, by computing ${\rm n}_{\rm U}$ in a thermal (Gibbs) state, we have that ${\rm n}_{\rm U} \xrightarrow{T\rightarrow0} {\rm Ch}$; hence, the generalization is faithful. Additionally, since $\Phi_{\rm U}$ is an observable, ${\rm n}_{\rm U}$ is also an observable. We will refer to this topological invariant ${\rm n}_{\rm U}$ as the \emph{Uhlmann number}.

In what follows, we classify the topological properties of some models of two-band topological insulators and superconductors at finite temperature according to the topological invariant ${\rm n}_{\rm U}$. We shall consider the analogous situation to \cite{Viyuela_et_al14} where the thermalization process preserves the number of particles and the crystalline momentum such as the equilibrium state splits in one-particle Gibbs states $\rho_{\bm{k}}^\beta={\rm e}^{-\beta H(\bm{k})}/Z$. Here, $H(\bm{k})$ is the one-particle Hamiltonian represented by a $2\times2$ matrix in the band indexes. Thus, the total Hamiltonian of these systems is a quadratic form $H_{\rm s}=\sum_{\bm k\in {\rm BZ}}\Psi_{\bm k}^{\dagger}H({\bm k})\Psi_{\bm k}$. In the case of insulators, $\Psi_{\bm k}=(a_{\bm k},b_{\bm k})^{\rm t}$, where $a_{\bm k}$ and $b_{\bm k}$ represent two species of fermions, while for superconductors, $\Psi_{\bm k}=(c_{\bm k},c^{\dagger}_{-\bm{k}})^{\rm t}$ is the Nambu spinor for paired fermions with opposite crystalline momentum \cite{LibroBernevig}. We take lattice spacing $a=1$ throughout the text.

\smallskip

\noindent {\it 4. 2D topological superconductor.---}
Let us consider the chiral \textit{p}-wave superconductor \cite{Volovik_99,Read_00,Ivanov_01,LibroBernevig}. This system can host vortices with non-Abelian anyonic statistics \cite{Ivanov_01} that are of great relevance in proposals for topological quantum computation \cite{Nayak_08}. The lattice Hamiltonian for this model is
\begin{align}
H=\sum_{ij}[&-t(c^{\dagger}_{i+1,j}c_{i,j}+c^{\dagger}_{i,j+1}c_{i,j})-\frac{1}{2}(\mu-4t)c^{\dagger}_{i,j}c_{i,j}+\nonumber\\
&+\Delta(c^{\dagger}_{i+1,j}c^{\dagger}_{i,j}+{\rm i}c^{\dagger}_{i,j+1}c^{\dagger}_{i,j})+\text{h.c.}],
\label{Hs_pwave}
\end{align}
where $\mu$ is the chemical potential, $t$ is the nearest-neighbour hopping and $\Delta$ is the superconductive pairing.

Without lost of generality, we fix $t=|\Delta|=1/2$. By means of a Bogoliubov transformation, we obtain the Hamiltonian in the Nambu spinor basis in momentum space,
\begin{align}
H({\bm k})=&-\{\sin{(k_y)}\sigma_x+\sin{(k_x)}\sigma_y \nonumber\\
&+[\mu-2+\cos{(k_x)}+\cos{(k_y)}]\sigma_z\},
\label{H_QWZ}
\end{align}
here, $\sigma_{x,y,z}$ are the three Pauli matrices.

At $T=0$, the different topological phases as classified by the Chern number, Eq. \eqref{ChernN}, are: ${\rm Ch}=1$ if $0<\mu<2$, ${\rm Ch}=-1$ if $2<\mu<4$ and ${\rm Ch}=0$ otherwise. For nontrivial regions ${\rm Ch}=\pm1$, the system presents chiral Majorana modes at the edges.

At finite temperature, the different topological phases as classified by the Uhlmann number, Eq. \eqref{nU}, are graphically represented Fig. \ref{Fig1}. The system displays nontrivial topological phases ${\rm n}_{\rm U}=\pm1$ even at nonzero temperature provided it is below a certain critical value $T_c$, where ${\rm n}_{\rm U}$ goes to zero. This critical temperature $T_c$ reaches the maximum value at the middle points $\mu=1$ and $\mu=3$ of the topological phases ${\rm Ch}=\pm1$ at $T=0$. These points are the ones with the highest value of the gap. As expected, in the limit of $T=0$ we recover the same topological diagram as given by the Chern number.

Thus, we see that thermal topological phase transitions are not a unique phenomenon of the 1D case \cite{Viyuela_et_al14} and they may be also found in two-dimensional systems.

\begin{figure}[t]
	\includegraphics[width=\columnwidth]{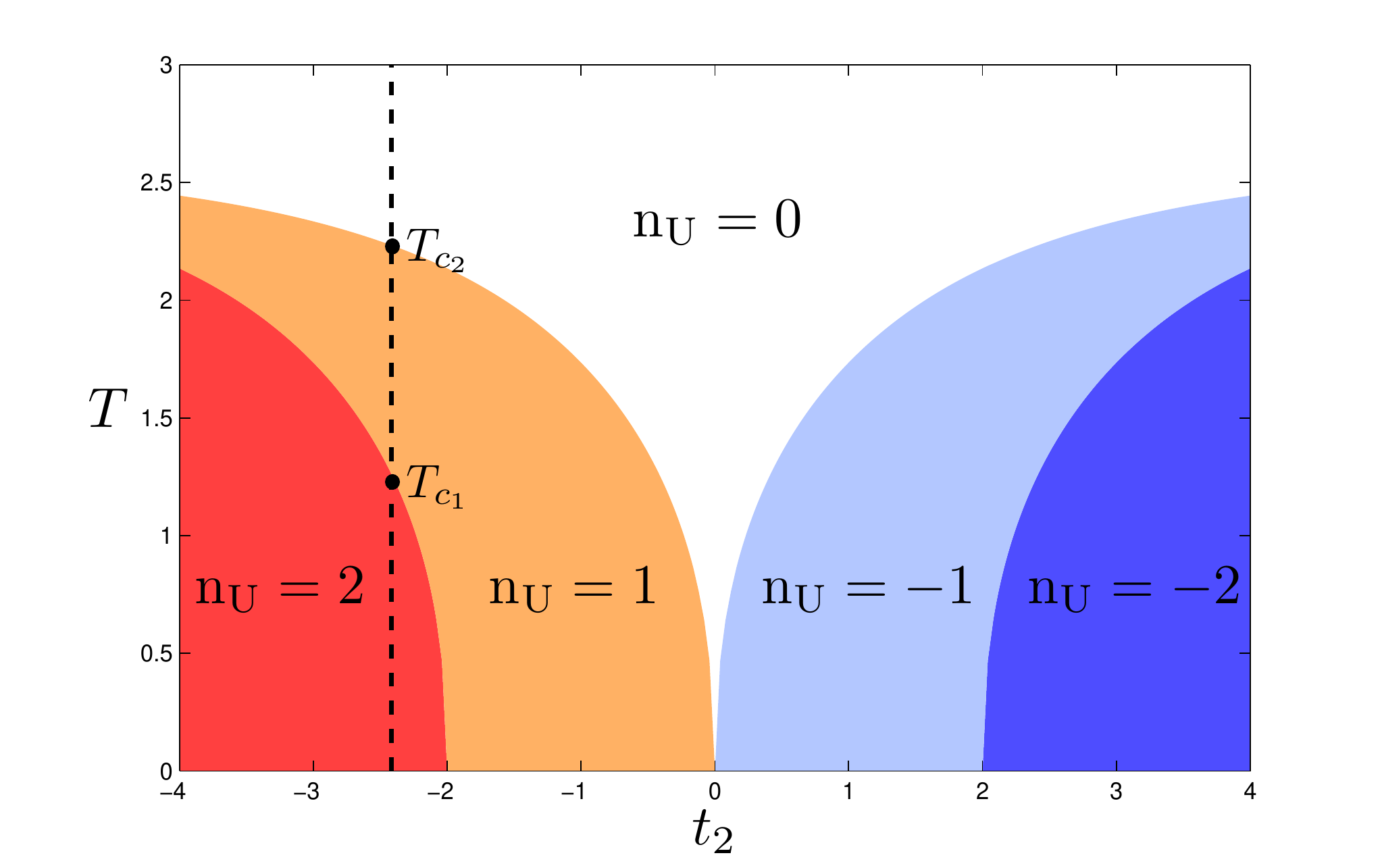}
	\caption{Uhlmann topological phase diagram for the model of Eq. \eqref{H_Simon}. The Uhlmann number is plotted for different values of $t_2$ and $T$. The dashed line highlights the purely thermal topological transitions between regimes of ${\rm n}_{\rm U}=2$, ${\rm n}_{\rm U}=1$ and ${\rm n}_{\rm U}=0$ by the sole effect of increasing $T$.}
	\label{Fig2}
\end{figure}

\smallskip

\noindent {\it 5. 2D topological insulator with high Chern number.---}
We consider here the model proposed in \cite{Sticlet_et_al12} that allows us to study 2D topological insulators with high values of the Chern number \cite{Yang_et_al12}, as in multi-band models like the Hofstadter model \cite{Hofstadter_76}, but still being two-band and analytically solvable.

This system is realized on a triangular lattice of fermionic atoms at each site with an internal orbital degree of freedom. The Hamiltonian is given by
\begin{align}
H=\sum_{ij}[&c^{\dagger}_{i+1,j}(t_1\sigma_x + {\rm i}t_3\sigma_z)c_{i,j}+c^{\dagger}_{i,j+1}(t_1\sigma_y + {\rm i}t_3\sigma_z)c_{i,j} \nonumber\\
&+c^{\dagger}_{i+1,j+1}(t_2\sigma_z)c_{i,j}+\text{h.c.}].
\label{Hs_r}
\end{align}

The Pauli matrices act on the orbital degrees of freedom at each site, which give rise to an orbital dependent nearest-neighbour hopping ($t_1,t_2,t_3$). In particular, the fermions can gain $\pi$ or $\pi/2$ phases depending on the initial and final orbital and position state when tunnelling. As in the Haldane model (see below), there is no net magnetic flux in the system, although time-reversal symmetry is broken.

Taking periodic boundary conditions the Hamiltonian in momentum space turns out to be
\begin{align}\label{H_Simon}
H({\bm k})&=2t_1\cos{(k_x)}\sigma_x+2t_1\cos{(k_y)}\sigma_y+\\
&+\{2t_2\cos{(k_x+k_y)}+2t_3[\sin{(k_x)}+\sin{(k_y)}]\}\sigma_z.\nonumber
\end{align}

Without loss of generality we take $t_1=t_3=1$. At zero temperature the topological phases as a function of $t_2$ are
\begin{equation}
{\rm Ch}= \begin{cases}
+2, & \text{if}~t_2<-2, \\
+1, & \text{if}~-2<t_2<0, \\
-1, & \text{if}~0<t_2<2, \\
-2, & \text{if}~t_2>2. \\
\end{cases}
\label{C_Simon}
\end{equation}
For $T\not=0$, the topological phase diagram according to Uhlmann number, Eq. \eqref{nU}, is shown in Fig. \ref{Fig2}. In this case we obtain two very remarkable and new effects with respect to the previous model. The first one relies on the existence of two critical temperatures $T_{c_1}$ and $T_{c_2}$. For instance, if at $T=0$ the system is in a topological phase with ${\rm n}_{\rm U}=2$, we observe that for $T<T_{c_1}$ the same phase is preserved. Then, for $T_{c_1}<T<T_{c_2}$ there is another thermal topological phase with ${\rm n}_{\rm U}=1$. If we now increase temperature even more, $T>T_{c_2}$ then the topological phase becomes trivial, ${\rm n}_{\rm U}=0$ (see Fig. \ref{Fig2}). Hence, there is a critical transition between phases with different (but nonzero) Uhlmann numbers by the sole effect of $T$. Thus, we have obtained a purely thermal transition between two different nontrivial topological regimes.

Secondly, at zero temperature we see in Eq. \eqref{C_Simon} that there are only nontrivial topological phases in this model. But, by increasing $T$, we can always end in a trivial phase with ${\rm n}_{\rm U}=0$. This supports the intuition that at sufficiently high temperatures the order is lost for any system.

\smallskip

\noindent {\it 6. Haldane model.---}
The Haldane model \cite{Haldane_88} was the first proposal of a 2D lattice of fermions without a constant magnetic field but with quantized Hall conductivity. It is a graphenelike model based on a honeycomb lattice with two different species of fermions (different sublattices), nearest-neighbors hopping $t_1$, next-nearest-neighbors hopping $t_2{\rm e}^{{\rm i}\phi}$, and a staggered potential $m$. For periodic boundary conditions the Haldane Hamiltonian in the reciprocal space is
\begin{align}
H({\bm k})&=\sum_{i}\{[2t_2\cos{\phi}\cos({\bm k}\cdot {\bm b}_i)]\mathds{1}+[t_1\cos({\bm k}\cdot{\bm a}_i)]\sigma_x+\nonumber\\
&+[t_1\sin({\bm k}_i\cdot {\bm a}_i)]\sigma_y+[m-2t_2\sin{\phi}\sin({\bm k}_i\cdot {\bm b}_i)]\sigma_z\},
\label{H_Haldane}
\end{align}
where ${\bm a}_i$ are the lattice vectors defining the Bravais lattice and ${\bm b}_i:={\bm a}_{i+1}-{\bm a}_{i-1}$. In particular, we take $t_1=4$ and $t_2=1$.

At $T=0$, the system presents topological order for $|m|<3\sqrt{3}|\sin{\phi}|$ with Chern number ${\rm Ch}=\pm1$ depending on the sign of $m$.

\begin{figure}[t]
	\includegraphics[width=\columnwidth]{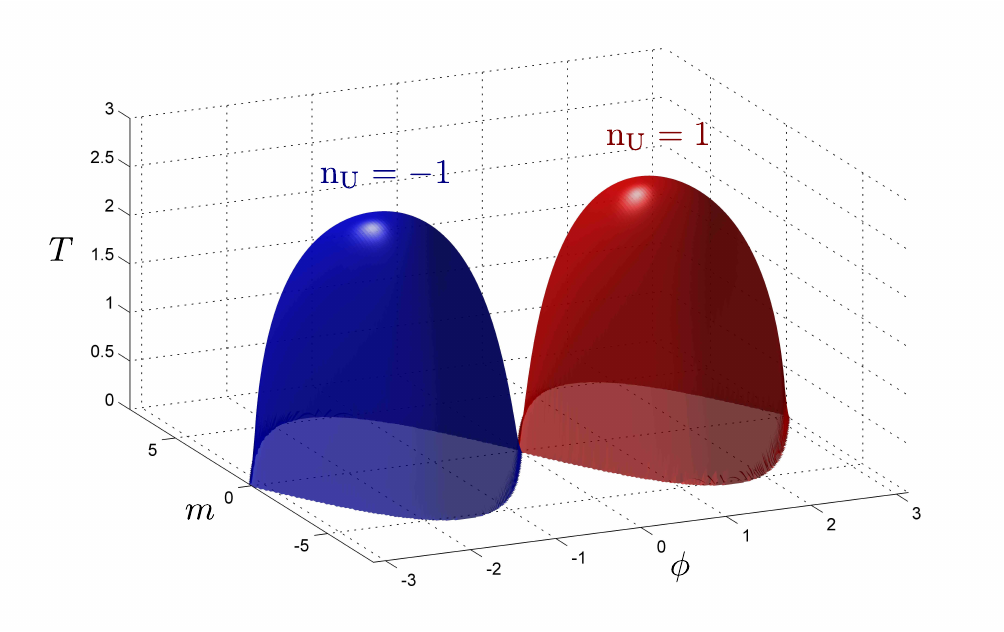}
	\caption{Uhlmann topological phase diagram for the Haldane model. Red color represents ${\rm n}_{\rm U}=1$ and blue ${\rm n}_{\rm U}=-1$. As we see, at $T=0$ the two well-known lobes of the Haldane model are obtained, and at a certain finite temperature $T_c$ the system goes to a trivial phase.}
	\label{Fig3}
\end{figure}

The topological phases at finite temperature as a function of $m$ and $\phi$ are depicted in Fig. \ref{Fig3}. The red and blue volumes represent ${\rm n}_{\rm U}=1$ and ${\rm n}_{\rm U}=-1$, respectively. Thus, an integer topological invariant $\pm 1$ is retained and a thermal topological phase is present up to some critical temperature $T_c$ where ${\rm n}_{\rm U}$ vanishes. Note that at $T=0$ we recover the well-known phase diagram for the Haldane model \cite{Haldane_88}.

Interestingly enough, the thermal topological properties of this model were first considered in \cite{Rivas_et_al13}. There, the topological indicator did not show a critical behavior with $T$ but shared the same pattern with $m$ and $\phi$ as ${\rm n}_{\rm U}$.

\smallskip

\noindent {\it 7. Conclusions and outlook.---} We have constructed a new topological invariant, the Uhlmann number ${\rm n}_{\rm U}$, that allows us to explore topological phases of fermion systems separated by purely thermal transitions. Notably, we find always a finite range of temperatures at which this topological order survives.

We remark that the existence of critical temperatures seems somehow natural in the Uhlmann approach. For thermal states, it sets on equal footing the temperature and the Hamiltonian parameters. Therefore, if there is a critical behavior as a function of tunnelings and/or staggered potentials, then certainly, one should obtain a critical behavior with temperature as well. Moreover, since by increasing $T$, the quantum coherence properties of any state are diminished, it is expected that the Uhlmann number decreases by warming the system up.

As explained in \cite{Viyuela_et_al14}, measurements of Uhlmann phases and numbers may be affordable by adapting experimental schemes such as \cite{Atala_et_al_12, Abanin_et_al_13, Demler_etal14}, that use interferometric setups for cold atoms in optical lattices. The mapping between Uhlmann amplitudes and pure state vectors in an enlarged Hilbert space should allow for measurements of the Uhlmann phase using an ancillary system.

Based on these results, we envision the possibility of extending the current classification of topological insulators and superconductors on several spatial dimensions \cite{Ludwig,Kitaev_2009} (also called the ``periodic table"), to the case of thermal topological states using the topological Uhlmann numbers introduced here.

\noindent {\em Note added:} While completing this work we were informed by Z. Huang
and D. Arovas about similar extensions of Uhlmann phases to 2D models \cite{Arovas14}.

\begin{acknowledgments}

 We thank the Spanish MINECO grant FIS2012-33152, FIS2009-10061, CAM research consortium QUITEMAD S2009-ESP-1594, European Commission
PICC: FP7 2007-2013 grant No.~249958, UCM-BS grant GICC-910758, FPU MEC grant and Residencia de Estudiantes.

\end{acknowledgments}


\begin{thebibliography}{99}



\bibitem{wenbook04}
X.-G. Wen. {\em Quantum Field Theory of Many-body Systems} (Oxford
University Press, Oxford, 2004).

\bibitem{wen89}
X.-G. Wen, Phys. Rev. B \textbf{40}, 7387 (1989).

\bibitem{wen_niu90}
X.-G. Wen and Q. Niu, Phys. Rev. B \textbf{41}, 9377 (1990).

\bibitem{ASW84}
D. Arovas, J. R. Schrieffer and F. Wilczek,
Phys. Rev. Lett. \textbf{53}, 722 (1984).

\bibitem{Haldane_88} F. D. M. Haldane, Phys. Rev. Lett. \textbf{61}, 2015 (1988).

\bibitem{kane_mele05}
C. L. Kane and E. J. Mele, Phys. Rev. Lett. \textbf{95}, 226801 (2005).

\bibitem{bernevig_zhang06}
B. A. Bernevig and S.-C. Zhang, Phys. Rev. Lett. \textbf{96}, 106802 (2006).

\bibitem{kane_mele05b}
C. L. Kane and E. J. Mele, Phys. Rev. Lett. \textbf{95}, 146802 (2005).

\bibitem{moore_balents07}
J. E. Moore and L. Balents, Phys. Rev. B \textbf{75}, 121306 (2007).

\bibitem{fu_kane_mele07}
L. Fu, C. L. Kane, and E. J. Mele, Phys. Rev. Lett. \textbf{98}, 106803 (2007);
X.-L. Qi, T. Hughes and S.-C. Zhang, Phys. Rev. B \textbf{78}, 195424 (2008).

\bibitem{rmp1}
M. Z. Hasan and C. L. Kane, Rev. Mod. Phys. \textbf{82}, 3045 (2010).

\bibitem{rmp2}
X.-L. Qi and S.-C. Zhang, Rev. Mod. Phys. \textbf{83}, 1057 (2011).

\bibitem{LibroBernevig}
B. A. Bernevig and T. L. Hughes, \textit{Topological Insulators and Topological Superconductors} (Princeton University Press, New Jersey, 2013).

\bibitem{VS13}
A. Vishwanath and T. Senthil, Phys. Rev. X \textbf{3}, 011016 (2013).

\bibitem{MKF13}
M. A. Metlitski, C. L. Kane and M. P. A. Fisher, Phys. Rev. B \textbf{88}, 035131 (2013).

\bibitem{Senthil14}
T. Senthil, arXiv:1405.4015.

\bibitem{YW13}
P. Ye and X.-G. Wen, Phys. Rev. B \textbf{89}, 045127 (2014).

\bibitem{wen_etal14}
Z.-X. Liu, Z.-C. Gu and X.-G. Wen, arXiv:1404.2818.

\bibitem{Bardyn_et_al13}
C.-E. Bardyn, M. A. Baranov, C. V. Kraus, E. Rico, A. \.{I}mamo\u{g}lu, P. Zoller and S. Diehl,  New J. Phys. \textbf{15}, 085001 (2013).

\bibitem{AFH}
R. Alicki, M. Fannes and M. Horodecki, J. Phys. A: Math. Theor. \textbf{42}, 065303 (2009).


\bibitem{Hastings}
M. B. Hastings, Phys. Rev. Lett. \textbf{107}, 210501 (2011).

\bibitem{Selfcorrecting}
Although stable topological phases for standard TOs can be established at high dimensions $D=4,6$:
R. Alicki, M. Horodecki, P. Horodecki and R. Horodecki, Open Syst. Inf. Dyn. \textbf{17}, 1 (2010); H. Bombin, R. W. Chhajlany, M. Horodecki and M. A. Martin-Delgado, New J. Phys. \textbf{15}, 055023 (2013).

\bibitem{Evert}
E. P. L. van Nieuwenburg and S. D. Huber, arXiv:1403.2387.

\bibitem{Viyuela_et_al12}
O. Viyuela, A. Rivas and M. A. Martin-Delgado, Phys. Rev. B \textbf{86}, 155140 (2012).

\bibitem{Viyuela_et_al14}
O. Viyuela, A. Rivas and M. A. Martin-Delgado, Phys. Rev. Lett. \textbf{112}, 130401 (2014).

\bibitem{Uhlmann86}
A. Uhlmann, Rep. Math. Phys. \textbf{24}, 229 (1986).

\bibitem{Uhlmann89}
A. Uhlmann, Ann. Phys. (Berlin) \textbf{501}, 63 (1989).

\bibitem{Hubner93}
M. H\"{u}bner, Phys. Lett. A \textbf{179}, 226 (1993).

\bibitem{Ericsson2003} M. Ericsson, A. K. Pati, E. Sj\"oqvist, J. Br\"annlund and D. K. L. Oi, Phys. Rev. Lett. \textbf{91}, 090405 (2003).

\bibitem{Sjoqvist_07} J. Aberg, D. Kult, E. Sj\"oqvist and D. K. L. Oi, Phys. Rev. A \textbf{75}, 032106 (2007).

\bibitem{Zhu_et_al11} J. Zhu,  M. Shi, V. Vedral, X. Peng, D. Suter and J. Du, EPL \textbf{94},  20007 (2011).

\bibitem{Eguchi} T. Eguchi, P. B. Gilkey and A. J. Hanson, Phys. Rep. \textbf{66}, 213 (1980).

\bibitem{Nakahara} M. Nakahara, \textit{Geometry, Topology and Physics} (CRC Press, Bristol, 2003).

\bibitem{footnote} This second argument also implies vanishing higher order Chern numbers for the Uhlmann connection.

\bibitem{Resta} R. Resta, Rev. Mod. Phys. \textbf{66}, 899 (1994).

\bibitem{Xiao} D. Xiao, M.-C. Chang and Q. Niu,  Rev. Mod. Phys. \textbf{82}, 1959 (2010).

\bibitem{Abanin_et_al_13}
D. A. Abanin, T. Kitagawa, I. Bloch and E. Demler, Phys. Rev. Lett. \textbf{110}, 165304 (2013).

\bibitem{Volovik_99}
G. E. Volovik, JETP Lett. \textbf{70}, 609 (1999).

\bibitem{Read_00}
N. Read and D. Green, Phys. Rev. B \textbf{61}, 10267 (2000).

\bibitem{Ivanov_01}
D. A. Ivanov, Phys. Rev. Lett. \textbf{86}, 268 (2001).

\bibitem{Nayak_08}
C. Nayak, S. H. Simon, A. Stern, M. Freedman and S. Das Sarma, Rev. Mod. Phys. \textbf{80}, 1083 (2008).

\bibitem{Sticlet_et_al12}
D. Sticlet, F. Pi\'echon, J.-N. Fuchs, P. Kalugin and P. Simon, Phys. Rev. B \textbf{85}, 165456 (2012).

\bibitem{Yang_et_al12} See also: S. Yang, Z.-C. Gu, K. Sun and S. Das Sarma, Phys. Rev. B \textbf{86}, 241112(R) (2012).

\bibitem{Hofstadter_76} D. R. Hofstadter, Phys. Rev. B \textbf{14}, 2239 (1976).

\bibitem{Rivas_et_al13}
A. Rivas, O. Viyuela and M. A. Martin-Delgado, Phys. Rev. B \textbf{88}, 155141 (2013).

\bibitem{Atala_et_al_12}
M. Atala, M. Aidelsburger, J. T. Barreiro, D. Abanin, T. Kitagawa, E. Demler and I. Bloch, Nat. Phys. \textbf{9}, 795 (2013).

\bibitem{Demler_etal14}
F. Grusdt,  D. Abanin, and E. Demler, Phys. Rev. A \textbf{89}, 043621 (2014).

\bibitem{Ludwig} A. P. Schnyder, S. Ryu, A. Furusaki and A. W. W. Ludwig, Phys. Rev. B \textbf{78}, 195125 (2008).

\bibitem{Kitaev_2009} A. Kitaev, AIP Conf. Proc. \textbf{1134}, 22 (2009).

\bibitem{Arovas14}
Z. Huang and D. P. Arovas, Phys. Rev. Lett. \textbf{113}, 076407 (2014).

\end{thebibliography}
\end{document}